\documentclass[a4paper,fleqn,usenatbib]{mnras}

\usepackage{amsopn}

\usepackage{color}
\usepackage{caption}
\usepackage{relsize}
\usepackage{graphicx}
\usepackage{amsmath}
\usepackage{amssymb}
\usepackage{pdflscape}
\usepackage{rotating}
\usepackage{longtable}
\usepackage{afterpage}

\title[Polarization \& ML for unidentified {\it Fermi} sources]{Probing the unidentified {\it Fermi} blazar-like population using optical polarization and machine learning}
\author[Liodakis, I.]
{I. Liodakis$^{1}$\thanks{ilioda@stanford.edu} \& D. Blinov$^{2,3,4}$\\
$^{1}$KIPAC, Stanford University, 452 Lomita Mall, Stanford, CA 94305, USA\\
$^{2}$Department of Physics and Institute for Theoretical and Computational Physics (ITCP), University of Crete, 71003, Heraklion, Greece\\
$^{3}$Foundation for Research and Technology - Hellas, IESL, Voutes, 7110 Heraklion, Greece\\
$^{4}$Astronomical Institute, St. Petersburg State University, Universitetsky pr. 28, Petrodvoretz, 198504 St. Petersburg, Russia\\
}

\begin{document}

\maketitle
\label{firstpage}
\begin{abstract}
The {\it Fermi} $\gamma$-ray space telescope has revolutionized our view of the $\gamma$-ray sky and the high energy processes in the Universe. While the number of known $\gamma$-ray emitters has increased by orders of magnitude since the launch of {\it Fermi}, there is an ever increasing number of, now more than a thousand, detected point sources whose low-energy counterpart is to this day unknown. To address this problem, we combined optical polarization measurements from the RoboPol survey as well as other discriminants of blazars from publicly available all-sky surveys in machine learning (random forest and logistic regression) frameworks that could be used to identify blazars in the {\it Fermi} unidentified fields with an accuracy of $>$95\%.  Out of the potential observational biases considered, blazar variability seems to have the most significant effect reducing the predictive power of the frameworks to $\sim80-85\%$. We apply our machine learning framework to six unidentified {\it Fermi} fields observed using the RoboPol polarimeter. We identified the same candidate source proposed by Mandarakas et al. for 3FGL~J0221.2+2518.
\end{abstract}


\section{Introduction}\label{introduc}

The {\it Fermi} $\gamma$-ray space telescope ({\it Fermi}) has detected more than 3000 point sources since its launch in 2008 \citep{Acero2015}. The galactic population has been dominated by pulsars, while the vast majority of the extragalactic population (accounting for 58\% of the total number of associated sources) consists of blazars: active galactic nuclei (AGN) with jets oriented close to our line of sight \citep{Blandford2018}. However, more than 1/3 of the total number of detected sources (as of the 3rd {\it Fermi} catalog) are yet unassociated with a low-energy counterpart. This is of course due to {\it Fermi}'s localization uncertainty which is of the order of a few arcminutes, much larger that e.g., a typical optical telescope. 

The continuing observations of {\it Fermi} seems to have two effects: 1) faint extragalactic $\gamma$-ray sources appear above the background (e.g., starburst galaxies, misaligned active galaxies); 2) there is an ever increasing number of unidentified sources. The latter is often a consequence of the limited amount of follow up observations which is not nearly as much as what is needed to associate even a fraction of the unidentified population.

Typically, an AGN is associated with a detected point source in the {\it Fermi} catalog using either the spatial coincidence of the {\it Fermi} point source with a source of a given catalog (Bayesian association method, \citealp{Abdo2010-V}) or the likelihood ratio association method which estimates an association probability based on the multiband (radio, optical, X-ray) fluxes of sources within {\it Fermi's} positional uncertainty \citep{Ackermann2015}. Point sources not associated with a counterpart using these methods are labeled as unidentified. Several other methods have been attempted to identify AGN in the unidentified {\it Fermi} fields (UFF). Popular methods include using the multiwavelength properties of the sources (e.g., \citealp{Acero2013}); the jet morphology revealed through Very Long Baseline Array (VLBI) observations \citep{Kovalev2009}; and infrared colors: blazars occupy a certain space in the infrared color-color diagram (e.g., \citealp{Massaro2011}). Most recently, \cite{Mandarakas2018} demonstrated that optical polarization can be a powerful probe for the $\gamma$-ray emitting AGN population. The method relies on the fact that $\gamma$-ray loud blazars have been found to be systematically more polarized than $\gamma$-ray quiet blazars \citep{Angelakis2016}. For sources above the galactic plane, where the interstellar dust induced polarization is expected to be less than 1-2\%, optical polarization could be used to confidently identify 544$\pm$10  blazars and blazar-like sources still hiding in the UFFs \citep{Mandarakas2018}.

Although each method has its limitations, a combination of different probes could prove to be an effective tool in the fight against the increasing number of unassociated $\gamma$-ray emitting sources. In this work, we combine optical polarization observations with publicly available data from various all-sky surveys in machine learning (ML) frameworks and explore the potential of such tools. In section \ref{samp} we present the sample and observables used in this analysis, in section \ref{machine} we built machine learning frameworks and evaluate their performance against observational biases, and in section \ref{application} we apply our framework to several unidentified fields of {\it Fermi} observed using the RoboPol polarimeter. We conclude in section \ref{concl}.

\section{Sample \& Analysis}\label{samp}

\begin{figure}
\resizebox{\hsize}{!}{\includegraphics[scale=1]{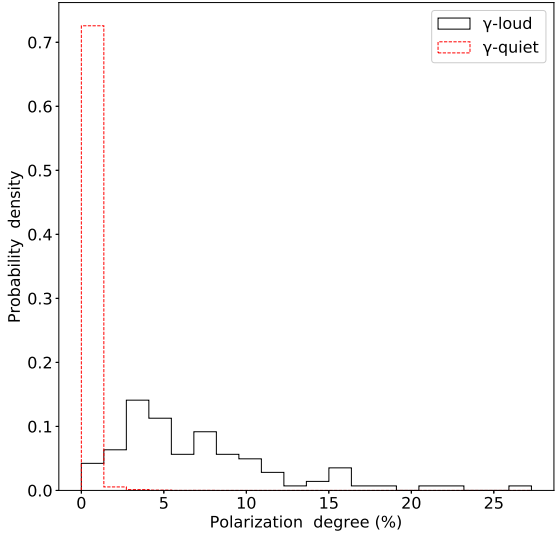} }
 \caption{Optical polarization degree distribution for the $\gamma$-ray loud and $\gamma$-ray quiet samples.}
\label{plt:polarization}
\end{figure}

\begin{figure}
\resizebox{\hsize}{!}{\includegraphics[scale=1]{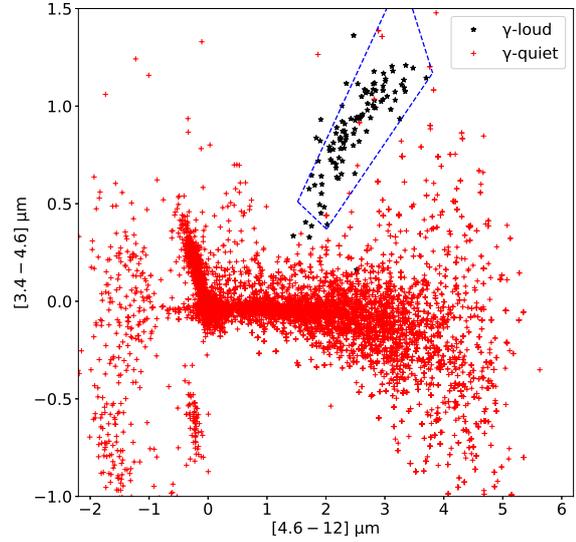} }
 \caption{WISE infrared colors (3.4-4.6$\rm \mu$m versus the 4.6-12$\rm \mu$m) for the $\gamma$-ray loud and $\gamma$-ray quiet samples. The blue dashed lines mark the blazar strip parametrized in \citealp{Massaro2012}.}
\label{plt:infrared_colors}
\end{figure}

\begin{figure}
\resizebox{\hsize}{!}{\includegraphics[scale=1]{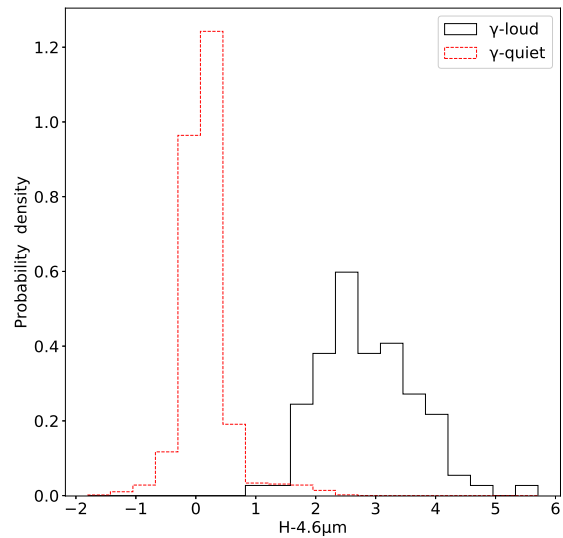} }
 \caption{H-band-4.6$\rm \mu$m distribution for the $\gamma$-ray loud and $\gamma$-ray quiet samples.}
\label{plt:hband}
\end{figure}

Our initial sample consists of $\gamma$-ray loud blazars observed in optical polarization by the RoboPol program\footnote{http://robopol.org/}. RoboPol observed a $\gamma$-ray flux-limited unbiased sample of blazars with high cadence for three years using a novel 4-channel polarimeter \citep{King2014}. The sample was defined using strict statistical criteria (for details on the sample selection see \citealp{Pavlidou2014}) making it suitable for statistical studies.  The main goal of the RoboPol project was to explore the polarization properties of the jets and provide insight on the nature of the electric vector position angle (EVPA) rotations seen in blazars \citep{Blinov2015,Blinov2016,Blinov2016-II,Blinov2018}. Apart from the main sample (62 sources) that was monitored during those three years, roughly 200 in total $\gamma$-ray loud blazars were observed at least once as part of the so-called ``June survey'' that was used for the final sample selection of RoboPol \citep{Pavlidou2014}. The sample observed as part of the June survey was also selected based on $\gamma$-ray flux ($\rm F(>100)~MeV > 2\times10^{-8}cm^{-2}s^{-1}$ in the second {\it Fermi} catalogue), optical magnitude ($\leq18$), sky position (Galactic latitude $|b|>10$), and elevation ($E>30$ for at least 30 minutes in June). From all observed blazars we selected those with at least three observations (93 sources). We complimented our sample with 11 additional $\gamma$-ray loud blazars observed by the Steward Observatory monitoring program\footnote{http://james.as.arizona.edu/$\sim$psmith/Fermi/} \citep{Smith2009}. The total number of $\gamma$-ray loud blazars in our sample is thus 104\footnote{Roughly 27\% of the sources are low synchrotron peaked, 20\% are intermediate synchrotron peaked, 17\% are high synchrotron peaked, and 36\% without synchrotron peak classification \citep{Abdo2010-II,Angelakis2016}.} (hereafter $\gamma$-ray loud (GL) sample).

The RoboPol polarimeter allows for the simultaneous measurement of the Stokes q and u vectors without any moving parts. Each source on the sky is projected four times (four spots in a cross-like pattern) on the CCD camera each having its polarization orientation rotated by 67.5$^o$. This allows for measurement of the polarization degree and EVPA of a source in a single exposure. For enhanced sensitivity and to block unwanted light, a mask is placed in the center of the CCD, the location where the source of interest is typically positioned. A consequence of the RoboPol design is that sources outside the mask could have one (or more) of their spots overlap with a neighboring source preventing us from confidently measuring either source's polarization parameters (see figure 6 in \citealp{Panopoulou2015}),  especially in crowded fields. For every blazar observed with the RoboPol polarimeter, we use the automated reduction pipeline to extract the polarization degree for the sources in the 13'$\times$13' arcminute field of view of the RoboPol instrument excluding sources affected by such overlap (for details on the RoboPol pipeline and reduction process see \citealp{King2014,Panopoulou2015}). This process yielded polarization measurements for 5837 sources. Additionally, we include 6379 sources from the \cite{Heiles2000} catalogue of polarized stars residing above and below the galactic plane ($|b|>10^o$ and  $|b|<-10^o$)  resulting in a total of 12216 sources (hereafter $\gamma$-ray quiet (GQ) sample). Figure \ref{plt:polarization} shows the median polarization degree of the GL and GQ samples. Clearly, the GL sources show systematically higher polarization due to the synchrotron nature of their emission as opposed to the, most likely, dust-induced polarization of the emission in GQ sources.

We use the Wide-field Infrared Survey Explorer (WISE) all-sky survey\footnote{http://wise2.ipac.caltech.edu/docs/release/allsky/} and the two micron all sky survey (2MASS, \citealp{Skrutskie2006} to obtain the 3.4, 4.6, 12$\mu$m, and H-band magnitudes for all the sources in the GL and GQ samples. The WISE colors have been successfully used in the past to find blazars among the unidentified {\it Fermi} sources (e.g., \citealp{Massaro2011}). Figure \ref{plt:infrared_colors} shows the WISE color-color diagram along with the so called ``blazar strip'' found in \cite{Massaro2011,Massaro2012}. While there is some minor overlap, the GL sample occupies a very distinct area in the infrared color-color diagram. We also estimate the H-band--4.6$\mu$m color. In \cite{Majewski2011} stars were shown to have a narrow distribution. The GQ sample (which is mostly comprised of stars) shows a similar narrow distribution whereas GL sources show a wider distribution shifted to higher values (Figure \ref{plt:hband}). In addition, stars and nearby galactic sources could have measurable proper motion while we do not expect any significant proper motion in distant extragalactic sources (i.e., blazars). We use the GAIA DR2 catalogue \citep{Gaia2016,Gaia2018} to estimate the proper motion (pm) and its uncertainty ($\sigma_{pm}$) for all the sources in the GL and GQ samples. If the source has a pm/$\sigma_{pm} <3$ we set pm=0. Otherwise we retain the estimated pm value. As expected, most of the sources in the GL sample have pm consistent with zero. There are 10 sources that show non-zero pm. This could be due to wrongful estimation of the pm (e.g., fitting for pm did not converge) or underestimated $\sigma_{pm}$. Whichever the case, we have verified that, for these 10 sources, using the estimated pm values or setting them equal to zero does not have any significant impact on the ML framework's predictive power.

\section{A machine learning approach}\label{machine}

\begin{figure*}
\resizebox{\hsize}{!}{\includegraphics[scale=1]{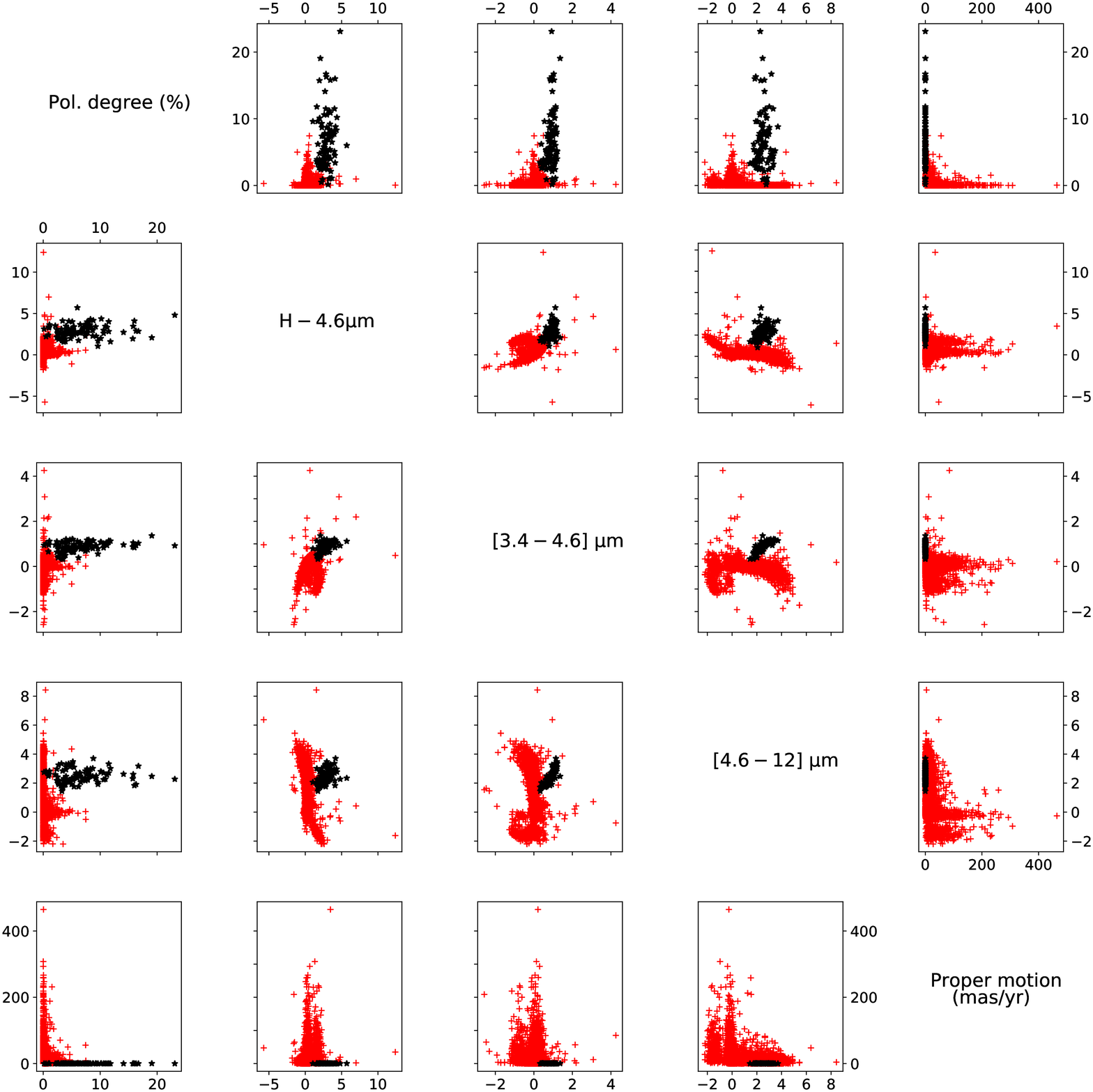} }
 \caption{Relation between two pairs of parameters. Black stars are for the $\gamma$-ray loud and red crosses for the $\gamma$-ray quiet samples.}
\label{plt:multiplot}
\end{figure*}

\begin{figure}
\resizebox{\hsize}{!}{\includegraphics[scale=1]{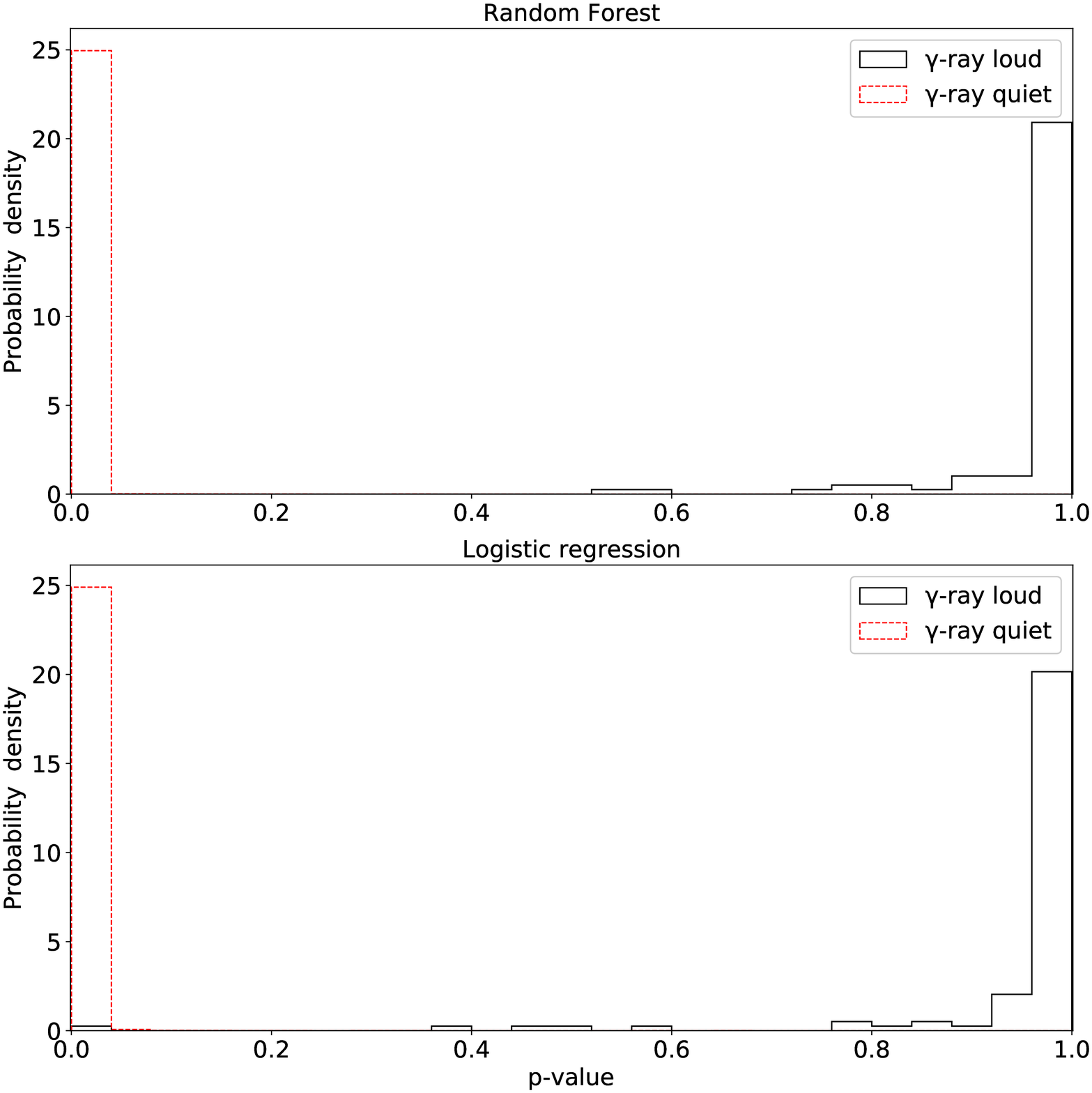} }
 \caption{Probability of emitting $\gamma$-rays for the GL and GQ sources in the test sample. The top panel is for the random forest while the bottom panel is for the logistic regression model.}
\label{plt:randforest_prob}
\end{figure}

The observables described above allow us to take advantage not only of the intrinsic properties of the $\gamma$-ray emitting jets but also the properties of the foreground sources and enable us to differentiate them. Figure \ref{plt:multiplot} shows the relationship between all pairs of observables for both GL and GQ samples. We first chose to combine them using a random forest model \citep{Breiman2001}. Random forests have been increasingly used in astronomy, and successfully used in classifying AGN sources of unknown type using their $\gamma$-ray properties (e.g., \citealp{Hassan2013,SazParkinson2016}). They are a method for classification by building a number of de-correlated decision trees with randomly drawn input variables from the training set. Each tree then casts a vote on the classification of the target source. The majority vote is the final result. Random forests are built on the principles of bootstrap aggregating (also known as bagging) that reduce the variance of the model achieving better classification results. We use the {\it sklearn} python implementation \citep{scikit-learn} to set up a random forest framework initially consisting of 20 trees using 80\% of the GL and GQ samples for training and 20\% for testing. Figure \ref{plt:randforest_prob} shows the probability of classifying a test source as GL for this particular choice for the number of decision trees and training sample size. We varied the number of trees from [20,200] with a 10 tree step as well as the percentage of sources in the testing sample between 5-50\% to evaluate the stability of the ML framework. In any combination of the above the accuracy of the framework remains within $\leq$2\% of the original setup ($>95\%$). To further assess the stability of the framework we perform two tests. First, we create simulated training and testing samples by randomly excising 30\% up to 70\% of the initial samples and repeat the analysis.  Even after $10^3$ trials the accuracy of the framework remained above 95\%. Second, we artificially misclassified sources randomly in the training sample starting from 1\% of the sample up to 100\%. We found that if no more than 20\% of the sample is misclassified the accuracy of the framework remains above 90\%.

An alternate approach which also serves as a validity check for the random forest is a logistic regression model. Logistic regression is a predictive regression method using a binary logistic model the outcome of which can be either ``success'' or ``fail''. The logistic model is defined as,
\begin{equation}
l=\beta_0 + \beta_1 X_1 + \beta_2 X_2 +...+ \beta_n X_n,
\end{equation} 
where $\beta_0$ is a constant value (intercept) and $[\beta_1,\beta_2,...,\beta_n]$ are the coefficients of the $[X_1,X_2,...,X_n]$ input variables. The probability for a given set of input parameters is given by,
\begin{equation}
p=\frac{1}{1+e^{-l}}.
\end{equation}
For $p>0.5$ (or alternatively if $l>0$) the outcome is ``success'' or in our case indicates a $\gamma$-ray emitting source. From the training procedure we obtain $\beta_0=-3.31$, and $[\beta_1,\beta_2,\beta_3,\beta_4,\beta_5]=[1.45,0.95,0.74,-0.29,-2.67]$. For our training $[X_1,X_2,X_3,X_4,X_5]=[\rm Pol.~degree~(\%),\rm ~3.4-4.6\mu{m},\rm ~ 4.6-12\mu{m},\rm ~H-4.6\mu{m},\rm ~pm]$. The logistic regression yielded an accuracy of 95\%  similar to the results of the random forest.

Although in both cases we achieve a high degree of accuracy, instrumental/observational systematics and/or the ever-present AGN variability can affect the measurements of the parameters described above hampering the accuracy of an already trained network. It is then important to assess the stability of the predictive power of our ML frameworks given the aforementioned effects.
\subsection{Dependence on individual parameters}\label{sec:dep_param}
\begin{figure}
\resizebox{\hsize}{!}{\includegraphics[scale=1]{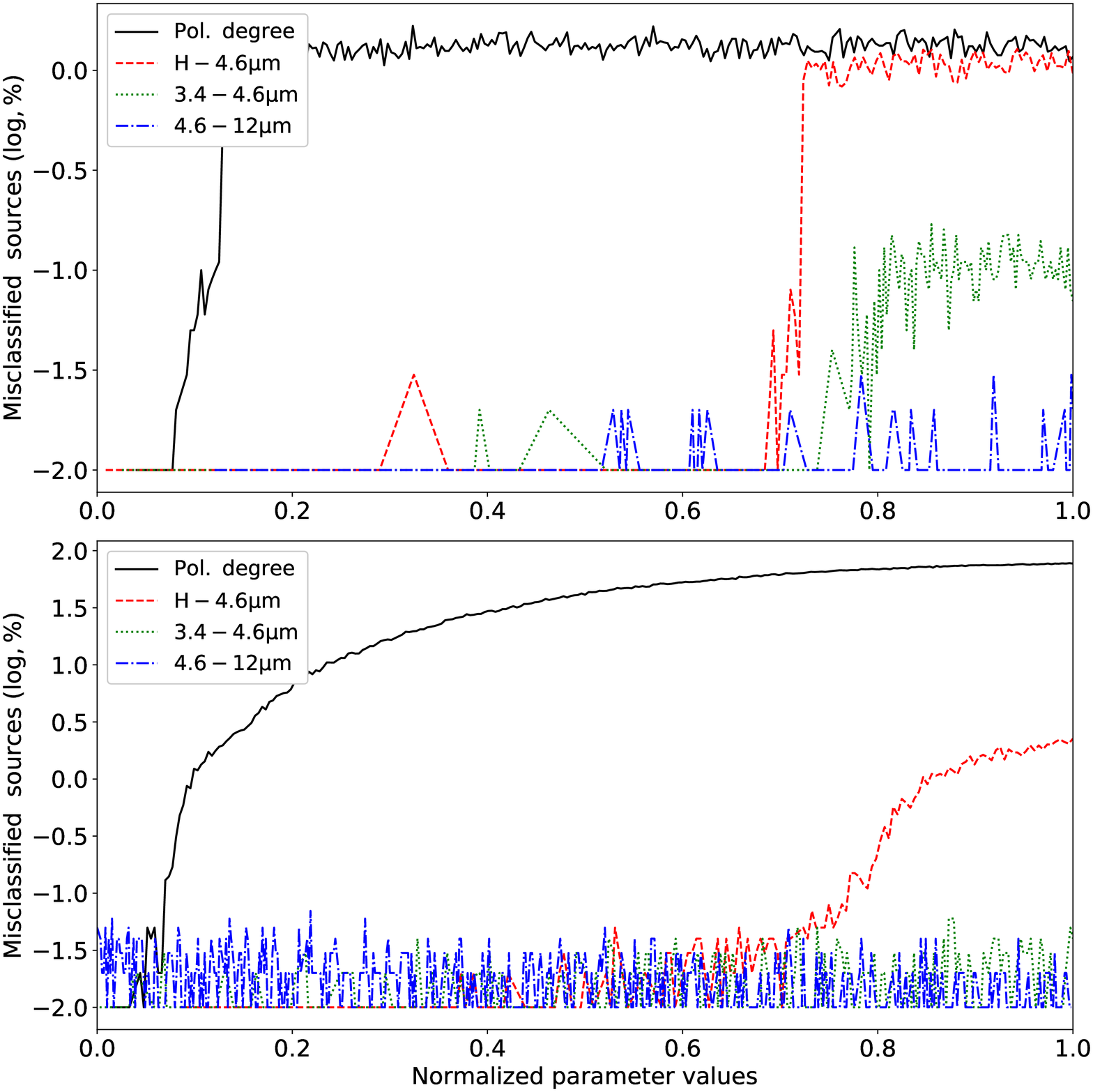} }
 \caption{Logarithm of the percentage of sources misclassified as $\gamma$-ray emitters by varying one variable at a time. The top panel is for the random forest while the bottom panel is for the logistic regression model.}
\label{plt:param_dep}
\end{figure}

Individual measurements of the source parameters could be affected due to a number of reasons. Although the dust-induced polarization is expected to drop significantly above the galactic plane, intrinsically polarized galactic sources (e.g., peculiar stars), abnormally dusty regions in a given line of sight, or even erroneous polarization measurements could result in non $\gamma$-ray sources having comparable levels of polarization with the GL sample. In addition, source confusion might affect crowded fields leading to erroneous color estimation. This effect might be particularly relevant for the infrared observations given the larger point spread function of the infrared telescopes compared to the optical. To assess the stability of the ML framework to these effects we create simulated test samples by drawing random values from the parameter distributions from the GQ sample for all parameters but one. Since we are evaluating the possibility of erroneous measurement or abnormally high values we do not expect correlation between observations by different instruments. The parameter under investigation takes values from the minimum of the GQ sample to the maximum value of the GL sample with a step of roughly 1\% of the target variable. We then use the already trained network and, after $10^4$ simulations, we estimate the fraction of incorrectly identified sources. We only use sources from the GQ sample for this exercise since we want to assess the fraction of sources that could be misclassified as GL source, and because intrinsic variability in GL sources can induce much larger variations in the parameters. We explore the effects of intrinsic variability in the subsection below. Figure \ref{plt:param_dep} shows the results of this exercise for both ML frameworks. We find that both frameworks are most sensitive to the polarization degree and $\rm H$-$4.6\mu{m}$ color. In both cases altering the $\rm H-4.6\mu{m}$ color results in a misclassification of only $\sim2\%$ at most. This is also true for the random forest and the polarization degree. However, the logistic regression shows a much higher dependence that could result in a misclassification of up to 80\%.

\subsection{Blazar variability}\label{bl_var}

Blazars are known to show violent and erratic variability throughout the electromagnetic spectrum and particularly in polarization (e.g., \citealp{Angelakis2016,Kiehlmann2017}). Given the large variations typically observed in the blazar light curves (e.g., \citealp{Liodakis2018-II,Liodakis2018}), it is possible for the parameter values of a GL source (except proper motion) to become comparable to those of a GQ source. It is then important to assess whether these variations can affect the classification of the GL sources. In order to take into account the intrinsic blazar variability we again create simulated GL samples and evaluate the performance of the ML frameworks. We use the values for the parameters of the beta distribution for the polarization degree found for the RoboPol blazars through a maximum likelihood approach in \cite{Angelakis2016} to draw a random value for each blazar. Then, we draw a random value for each of the infrared magnitudes from a Gaussian distribution assuming a 20\% spread and recalculate the infrared colors. This results in a spread of 2-3$^m$ typical for blazars. Once we have a simulated sample we use the framework to predict whether the simulated sources are $\gamma$-ray emitters. We repeat this process $10^4$ and estimate the percentage of sources that where not correctly identified. The logistic regression seems to be less affected by variability with a median value of 15$\pm3.5$\% misclassified sources whereas random forest shows a median of 22$\pm4$\%. Although the relation between the polarization and flux variations in uncertain, blazars are known to flare simultaneously in multiple bands (e.g., \citealp{Liodakis2017,Liodakis2018-II},). We repeat the above exercise twice. The first time we assume correlated variability between infrared colors but the polarization degree varies randomly (correlated infrared (CI) case), while the second time all parameters are correlated (correlated infrared and polarization (CIP) case). In both cases the random forest framework yields similar results (20$\pm4$\% for the CI and  19$\pm4$\% for the CIP case). The same is true for the logistic regression (roughly 12$\pm3$\% in both cases). Although increasing the spread of the parameters seems to increase the percentage of misclassified sources in all cases, for large values for the spread ($>100\%$) there is a saturation at $\sim35\%$ for the random forest and $\sim45\%$ for the logistic regression. However, such extreme cases should be very rare given the fact that the most bright and variable blazars emitting $\gamma$-rays have most likely already been identified.

\section{Application to unidentified {\it Fermi} sources}\label{application}

\begin{figure}
\resizebox{\hsize}{!}{\includegraphics[scale=1]{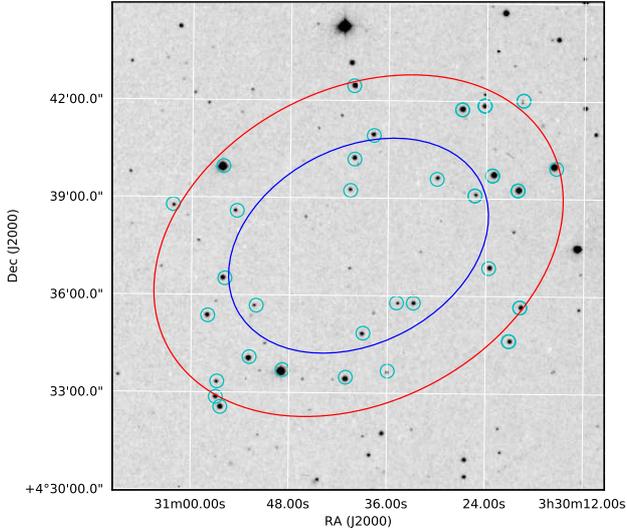} }
 \caption{Sources observed by RoboPol (small circles) within the positional accuracy (large circles, 68\% blue, 95\% red) of 3FGLJ0330.6+0437.}
\label{plt:UFF_field}
\end{figure}
\begin{figure}
\resizebox{\hsize}{!}{\includegraphics[scale=1]{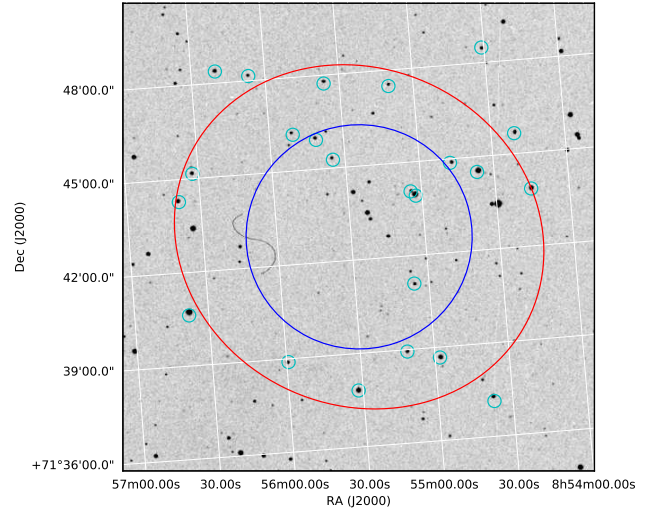} }
 \caption{Sources observed by RoboPol (small circles) within the positional accuracy (large circles, 68\% blue, 95\% red) of 3FGL~J0855.4+7142.}
\label{plt:UFF_field2}
\end{figure}

\begin{figure}
\resizebox{\hsize}{!}{\includegraphics[scale=1]{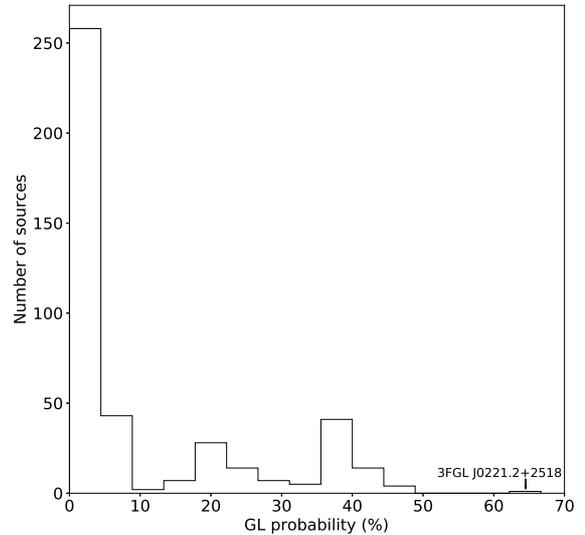} }
 \caption{Probability of a source being a $\gamma$-ray emitter for all the sources in the UFFs observed by RoboPol. The source identified as a potential $\gamma$-ray emitter in 3FGL~J0221.2+2518 is marked with an arrow.}
\label{plt:UFF_proba}
\end{figure}

\begin{figure}
\resizebox{\hsize}{!}{\includegraphics[scale=1]{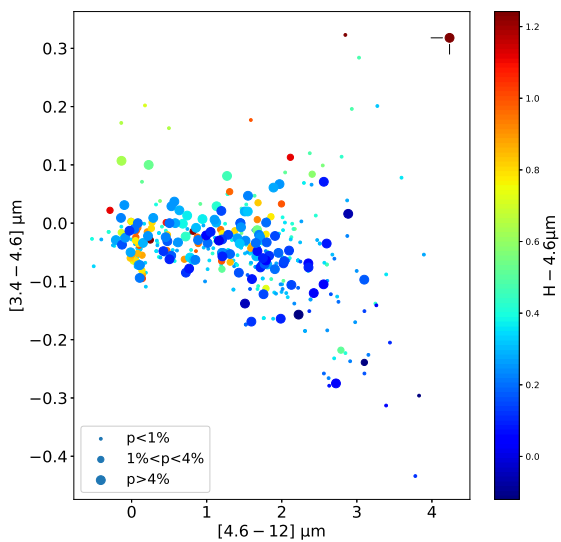} }
 \caption{Infrared colors for the sources in the six UFFs observed by Robopol. The size of the circles denoted the percentage of polarization. The candidate found in \citealp{Mandarakas2018} in 3FGL~J0221.2+2518 is marked with the vertical and horizontal lines.}
\label{plt:UFF_colors}
\end{figure}

Given the high fraction of misclassifications by the logistic regression framework found in section \ref{sec:dep_param}, for the application of our methodology we opted to use the random forest framework. \cite{Mandarakas2018} used optical polarization to identify potential candidates for the $\gamma$-ray emitters in 4 UFFs. Several sources were observed within the 3$\sigma$ positional uncertainty of each field ($\sim4$ arcminutes, \citealp{Acero2015}). Their analysis yield a candidate source in the 3FGL~J0221.2+2518 field at $\rm RA=02h21m33.3s$, $\rm DEC=+25\degr12\arcmin47.3\arcsec$.  Follow up spectroscopic observations confirmed its extragalactic origin while the comparison of the line ratios suggest a hybrid AGN-starburst galaxy. The source had not been included in published AGN catalogues and showed atypical, for a $\gamma$-ray emitting AGN, [3.4-4.6]$\mu$m and [4.6-12]$\mu$m colors. The three other UFFs considered in \cite{Mandarakas2018} were 3FGL~J0336.1+7500, 3FGL~J0419.1+6636, and 3FGL~J1848.6+3232. In our study we include two additional fields also observed by RoboPol: 3FGL~J0330.6+0437 and 3FGL~J0855.4+7142. Each field was covered in three exposures. The first one was centered at the reported position of each UFF and the other two exposures were offset by 1 arcminute.  Figures \ref{plt:UFF_field} and \ref{plt:UFF_field2} show the positional uncertainty of {\it Fermi} (68\% and 95\% uncertainty, large circles) overlaid on an SDSS image \citep{SDSS2018}. The small circles mark the sources observed by RoboPol.

We use the polarization estimates for all the sources observed in the UFFs and follow the same procedure as in section \ref{samp} to estimate their infrared colors and proper motion. Figure \ref{plt:UFF_proba} shows the random forest probability of a source emitting $\gamma$-rays for all the sources observed by RoboPol. Only one source is classified as a GL source in 3FGL~J0221.2+2518 with a 66\% probability marked with an arrow. The same source was proposed as the potential $\gamma$-ray emitter in \cite{Mandarakas2018}. No additional candidates were identified in any of the UFFs. However, that does not necessarily suggest that there are no blazar-like sources within those fields.  Given the wide field limitations of the RoboPol instrument (\citealp{Panopoulou2015}, see also section \ref{samp}) it is likely that the source was either not observed or source overlap and/or other systematics prevented us from measuring its polarization degree (Figure \ref{plt:UFF_field}, \ref{plt:UFF_field2}). In addition, the limited exposure time used for the observations ($\sim$30-40 minutes per exposure) allows us to confidently estimate the polarization degree only for sources brighter than $\sim18^m$ in the R-band. This prevents us from probing fainter sources that could be responsible for the $\gamma$-ray emission. 

Based on the discussion in section \ref{bl_var} it is also necessary to evaluate the probability of misclassifying blazar-like sources as non-$\gamma$-ray emitters. We use the ML probabilities of the simulated misclassified blazars in section \ref{bl_var} to estimate the number of blazars with p-value between 0.1 (GQ sources show $<0.1$ p-values in the test sample, figure \ref{plt:randforest_prob}) and 0.46 (the highest probability in the observed sample below the 0.5 threshold). We find that the majority of simulated misclassified blazars ($\sim$60\%) lay within that range suggesting that it is not unlikely for a misclassified blazar due to variability to exist within the observed fields. Additional observations are necessary to definitely exclude the possibility of a non blazar-like source responsible for the $\gamma$-ray emission in those UFFs. Figure \ref{plt:UFF_colors} shows the infrared colors and polarization of all the sources observed in all six UFFs. The source in the top right corner marked with the vertical and horizontal lines, clearly standing out from the majority of the sources, is the candidate source found by the ML framework. In this particular UFF, for the next best candidate the random forest yields a p-value of 0.3. Only 19\% of the simulated misclassified blazars have probabilities between $0.1<$p-value$<0.3$.

\section{Conclusions}\label{concl}

We have demonstrated that optical polarization combined with other discriminants of jets and foreground sources in machine learning frameworks can be a powerful tool in the search for the missing blazar/blazar-like population in the still unidentified fields of {\it Fermi}. We introduced two new observables as predictors: the $\rm H-4.6\mu{m}$ color and proper motion as measured by GAIA. Those allow us to effectively segregate nearby galactic sources from the extragalactic population where polarization can then easily distinguish synchrotron emission from e.g., dust induced polarization or other mechanisms producing low degrees of optical polarization. Our approach is particularly useful for two main reasons. First, the majority of data used in this work are from publicly available, easily accessible, all-sky surveys. In addition, several upcoming polarization surveys (e.g., PASIPHAE, \citealp{Tassis2018}) will cover a large fraction of the sky providing measurements for the majority of sources in the UFFs. All of the above makes our proposed methodology among the least observationally expensive methods to find candidate extragalactic $\gamma$-ray emitters. Second, while individual observational probes have their limitations (e.g., the source in 3FGL~J0221.2+2518 would not have been classified as a $\gamma$-ray emitter using the WISE colors), the combination of the observables discussed here limits their individual imperfections in identifying candidates. It should be noted that although 90\% of the GL sample is from the statistically complete unbiased RoboPol sample that describes the $\gamma$-ray emitting blazar population, the low number of sources used for the training of the framework could potentially affect its accuracy. Demographics could also be important, although there is no significant difference in the polarization degree \citep{Angelakis2016} or proper motion between populations, and the blazars occupy distinct regions in the infrared color space from other types of sources (e.g., \citealp{Massaro2011}). Additional measurements of $\gamma$-ray blazars (with an equally mixed composition of synchrotron peaked sources) would undoubtedly benefit our ML approach.

While confusion, erroneous measurements or other potential observational biases seem to have a small effect on the predictive power of the proposed ML frameworks ($<2\%$ of sources could be misclassified for the random forest framework when varying a given parameter), blazar variability seems to affect it significantly (up to $\sim22\%$). Multiple measurements, additional predictors not affected (or less affected) by variability, or alternative ML approaches could help reduce this effect.

Applying our method to UFFs, our best candidate for 3FGL~J0221.2+2518 is the same as the one proposed by \cite{Mandarakas2018}. No additional candidate was found in 5 other fields without, however, excluding the possibility that the $\gamma$-ray emitter was not among the sources observed in polarization. Instruments built for wide field polarimetry such as WALOP\footnote{http://pasiphae.science/walop} would remove this limitation by providing measurements for all the sources within the positional uncertainty of {\it Fermi} and could be used to effectively identify the $\gamma$-ray emitting sources in the UFFs. As noted in \cite{Mandarakas2018} there are two additional candidates for 3FGL~J0221.2+2518: a radio galaxy NVSS~J022126+251436 and a radio source detected by \cite{Schinzel2017}. Although our approach disfavors NVSS~J022126+251436 as the $\gamma$-ray emitter, we can not exclude the possibility of that radio source emitting $\gamma$-rays, given that it was not detected during the polarization survey of 3FGL~J0221.2+2518.

Our approach is in principle limited to blazars and blazar-like sources. However, application of the framework to the {\it Fermi} fields observed by \cite{Mandarakas2018} revealed that it can also be used to probe the non-traditional $\gamma$-ray AGN population. Polarization measurements of the increasing number of misaligned AGN detected by {\it Fermi} could allow us to build a more robust framework suitable for the entire extragalactic population of $\gamma$-ray emitters. Current and future large scale surveys (e.g., eROSITA, LSST) could also help identify additional observables that could be easily incorporated into the proposed ML frameworks providing even more robust identifications for the entirety of the AGN $\gamma$-ray emitting population.

\section*{Acknowledgments}
We thank G. Panopoulou and the referee T. Hassan for comments and suggestions that helped improve this work. DB acknowledges support from the European Research Council (ERC) under the European Union’s Horizon 2020 research and innovation programme under grant agreement No 771282. RoboPol is a collaboration involving the University of Crete, the Foundation for Research and Technology – Hellas, the California Institute of Technology, the Max-Planck Institute for Radioastronomy, the Nicolaus Copernicus University, and the Inter-University Centre for Astronomy and Astrophysics. Data from the Steward Observatory spectropolarimetric monitoring project were used. This program is supported by Fermi Guest Investigator grants NNX08AW56G, NNX09AU10G, NNX12AO93G, and NNX15AU81G. This publication makes use of data products from the Wide-field Infrared Survey Explorer, which is a joint project of the University of California, Los Angeles, and the Jet Propulsion Laboratory/California Institute of Technology, funded by the National Aeronautics and Space Administration. This publication makes use of data products from the Two Micron All Sky Survey, which is a joint project of the University of Massachusetts and the Infrared Processing and Analysis Center/California Institute of Technology, funded by the National Aeronautics and Space Administration and the National Science Foundation. This work has made use of data from the European Space Agency (ESA) mission
{\it Gaia} (\url{https://www.cosmos.esa.int/gaia}), processed by the {\it Gaia}
Data Processing and Analysis Consortium (DPAC,
\url{https://www.cosmos.esa.int/web/gaia/dpac/consortium}). Funding for the DPAC
has been provided by national institutions, in particular the institutions
participating in the {\it Gaia} Multilateral Agreement. Funding for the Sloan Digital Sky Survey IV has been provided by the Alfred P. Sloan Foundation, the U.S. Department of Energy Office of Science, and the Participating Institutions. SDSS-IV acknowledges
support and resources from the Center for High-Performance Computing at
the University of Utah. The SDSS web site is www.sdss.org. SDSS-IV is managed by the Astrophysical Research Consortium for the 
Participating Institutions of the SDSS Collaboration including the 
Brazilian Participation Group, the Carnegie Institution for Science, 
Carnegie Mellon University, the Chilean Participation Group, the French Participation Group, Harvard-Smithsonian Center for Astrophysics, 
Instituto de Astrof\'isica de Canarias, The Johns Hopkins University, 
Kavli Institute for the Physics and Mathematics of the Universe (IPMU) / 
University of Tokyo, the Korean Participation Group, Lawrence Berkeley National Laboratory, 
Leibniz Institut f\"ur Astrophysik Potsdam (AIP),  
Max-Planck-Institut f\"ur Astronomie (MPIA Heidelberg), 
Max-Planck-Institut f\"ur Astrophysik (MPA Garching), 
Max-Planck-Institut f\"ur Extraterrestrische Physik (MPE), 
National Astronomical Observatories of China, New Mexico State University, 
New York University, University of Notre Dame, 
Observat\'ario Nacional / MCTI, The Ohio State University, 
Pennsylvania State University, Shanghai Astronomical Observatory, 
United Kingdom Participation Group,
Universidad Nacional Aut\'onoma de M\'exico, University of Arizona, 
University of Colorado Boulder, University of Oxford, University of Portsmouth, 
University of Utah, University of Virginia, University of Washington, University of Wisconsin, 
Vanderbilt University, and Yale University.

\bibliographystyle{mnras}
\bibliography{bibliography} 

\label{lastpage}
\end{document}